# ZEPLIN IV: A 1-Ton Very Sensitive ZEPLIN II Extension for SUSY Dark Matter*

David B. Cline, Hanguo Wang, Y. Seo

*Department of Physics & Astronomy, University of California, Los Angeles, CA 90095 USA*

We present a concept of a one ton two-phase liquid Xenon detector based on the concept of the ZEPLIN II detector currently under construction by a UCLA/Torino/UKDM team. The ZEPLIN II detector may be installed in the Boulby laboratory early in 2002. The one ton detector design will benefit from the initial operations of ZEPLIN II.

## 1. Status of the Search For WIMPs

The current status of the search for WIMPs is confused; a recent summary can be found in Ref. 1:

1. The level is at about 1/2 event/kg/d;
2. The DAMA group is making strong claims for a discovery;
3. The CDMS group has shown that their data and those of DAMA are incompatible to 99.5% confidence level against the observation of WIMPs.

Theory, which allows WIMP rates from about $10^{-1}$ to $10^{-4}$ events/kg/d, gives poor guidance. We believe that it is essential that our method (a) resolve the CDMS/DAMA claims, and (b) cover the entire region of $10^{-1}$ to $10^{-4}$ events/kg/d.

## 2. Properties of Liquid Xenon and Development of a Xenon WIMP Detector

The key properties of liquid Xe are given in Table 1, and Fig. 1 shows the key method of discrimination.[2,3,4]

Starting in the early 1990s, the UCLA/Torino ICARUS group initiated the study of liquid Xe as a WIMP detector with powerful discrimination. Our most recent effort is the development of the two-phase detector. Figure 2 shows our 1-kg, two-phase detector and the principle of its operation. WIMP interactions are clearly discriminated from all important background by the amount of free electrons that are drifted out of the detector into the gas phase where amplification occurs. In Fig. 3, we show the resulting separation between backgrounds and simulated WIMP interactions (by neutron interaction). It is obvious from this plot that the discrimination is very powerful.

## 3. ZEPLIN II Design and Construction

After the success of the 1-kg, two-phase detector, two directions are being taken:

(1) Construction of a large two-phase detector to search for WIMPs. The UCLA/Torino group has formed a collaboration with the UK Dark Matter team to construct a 40-kg detector (ZEPLIN II) for the Boulby Mine underground laboratory (Fig. 4).
(2) Continuation of the R&D effort with liquid Xe to attempt to amplify the very weak WIMP signal. The second idea to test is inserting a CsI internal photo cathode to convert UV photons to electrons that are subsequently amplified by the gas phase of the detector. In Fig. 4b we show the latest design of the ZEPLIN II detector. The expected sensitivity of ZEPLIN II is shown in Figure 6.

---

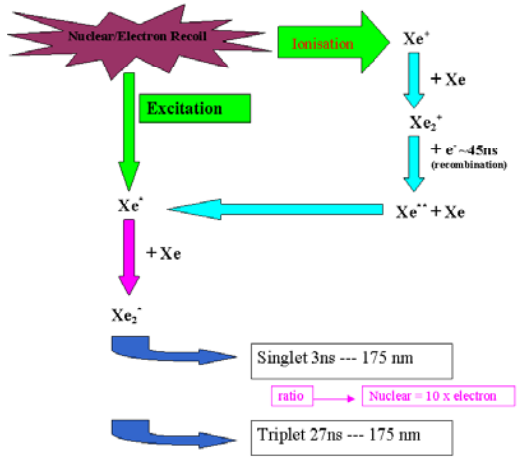

**Liquid Xenon Scintillation Mechanism**

Figure 1. Basic mechanism for the signal and background detection in liquid Xe.

**TABLE I. Liquid Xenon as a WIMP Detector**

1. Large mass available - up to tons.

   - Atomic mass: 131.29
     - Density: 3.057 gm/cm$^3$
     - $W_i$ value (eV/pair) 15.6 eV
     - No long-lived isotopes of xenon

2. Drift velocity: 1.7 mm/µ.s @ 250V/cm field

   - Decay time: 2 ns → 27 ns

3. Light yield > NaI, but intrinsic scintillator (no doping)

   ⇒ Excimer process very well understood

   ⇒ First excimer laser was liquid xenon in 1970!

The ZEPLIN II detector construction is started at UCLA and it will be completed by this fall. The central detector consists of:

1. PTFE cone which shapes the active liquid xenon zone,
2. Three liquid level meter to monitor the liquid level,
3. Ten copper rings to shape the static electric field for free ionization electrons to drift up to the gas phase,
4. Two wire frames to form a electron extraction field at the liquid and gas phase surface and to form a electro-luminescence field just above the surface,
5. Seven PMTs above the liquid to collect scintillation and luminescence photons.
6. Copper cast vessel to hold all the above in liquid xenon temperature,
7. Copper cast vacuum vessel to thermally insulate all the above,
8. Three 25kV HV feed-through to provide the power for the drift and luminescence field,
9. other signal, temperature sensing, level meter signal feed-throughs

Item 5 (PMTS) will be provided by Torino and everything else will be make at UCLA. The manufacturing is expected to finish this fall, and an initial assembly at UCLA is expected at the end of the fall and the RAL members will be present at the time to see the whole procedure so to ease the testing at RAL and installation at Boulby mine.

The copper casting is used because:

1) there will be no welding needed for the entire vessel system. This will guarantee the vacuum, and an extremely clean system since there are no dead surfaces on the inner vessel,

2) It is extremely cheap.

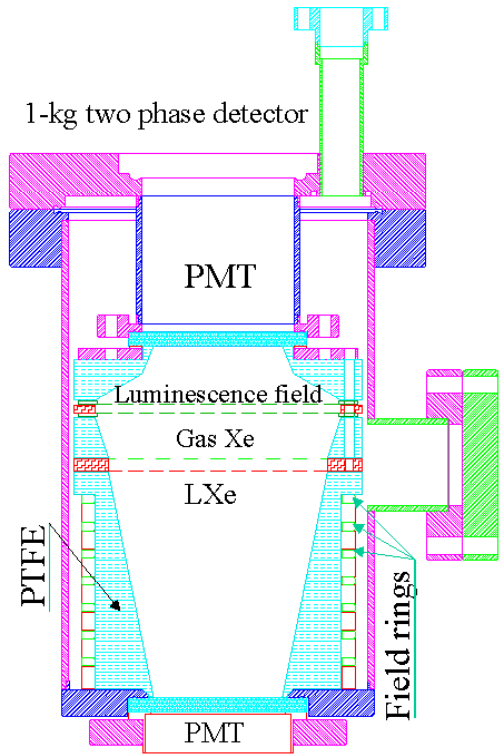
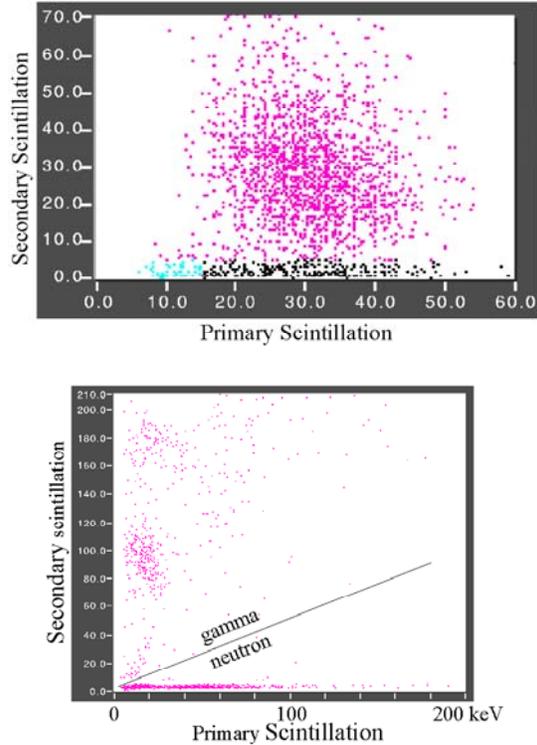

Figure 2. ZEPLIN II: Electroluminescence in gas (principle of a two-phase, 1-kg detector, developed by UCLA-CERN-Torino).

Figure 3. Secondary vs. primary scintillation plot in pure liquid Xe with mixed gamma-ray and neutron sources. The secondary scintillation are produced by proportional scintillation process in liquid Xe (top) and electroluminescent process in gaseous Xe (bottom).

### 4. Conceptual Design of a 1 Ton Detector: ZEPLIN IV

To consider a 1-ton detector (ZEPLIN IV) we adopt the design principle of the ZEPLIN II detector under construction now. As we complete and operate ZEPLIN II, adjustments to the design of ZEPLIN IV will be carried out by this team. The basic concept of ZEPLIN IV is shown in Figure 5.

- a) The PMTs are placed above the liquid and gas (two phase) system.[2,3,4.]
- b) Great care is taken to reduce any dead regions in the detector (just like ZEPLIN II).
- c) The engineering design principles will be the same as for ZEPLIN II.
- d) The H.V. system will be raised to the level to detect ionization if needed (from the primary).
- e) A possible CsI insert will be considered for a possible signal amplification.

We attempt to estimate the sensitivity of ZEPLIN IV in Figure 6—if correct, this would make ZEPLIN IV the most sensitive WIMP detector currently being studied. The dark matter signal levels expected in References 5 and 6 for example could ultimately require the sensitivity in Figure 6 and the construction of ZEPLIN IV.

We wish to thank M. Atac, P. Picchi, P. Smith, N. Smith and R. Preece for help.

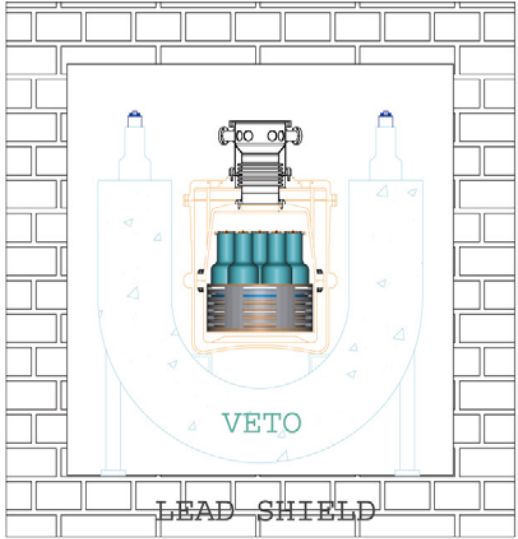 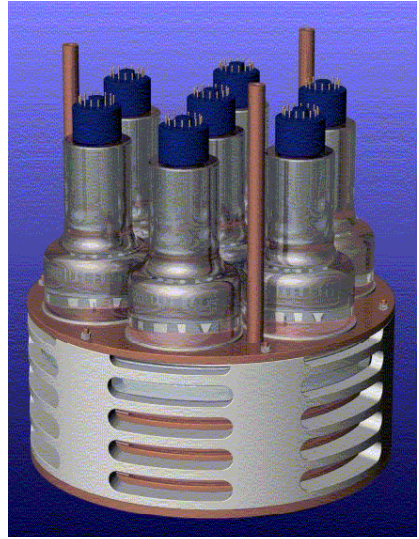

Figure 4. System setup for Xe target (40 kg total): (a) overall set up and (b) ZEPLIN II central detector.

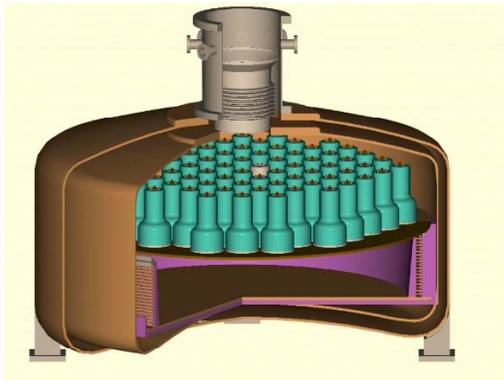

Figure 5. 1-ton scaled up (Zeplin IV) detector.

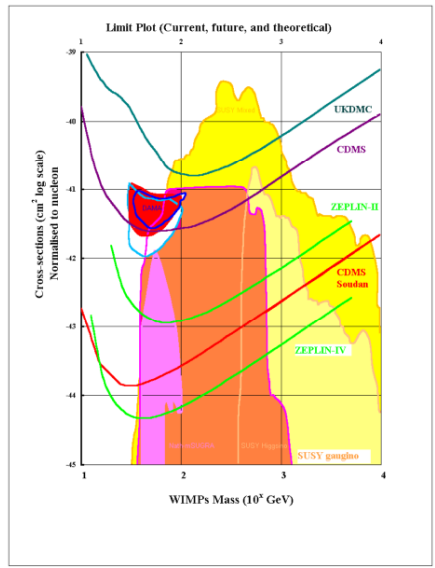

Figure 6. Limit plot (current, future and theoretical)

**References**

1. Sources of Dark Matter and Dark Energy in the Universe (Proc., 4th Int. Symp., Marina del Rey, CA, Feb. 2000) edited by David B. Cline (Springer Verlag, Heidelberg, 2001).
2. Cline, D., Curioni, A., Lamarina, A., et al., Astropart. Phys. 12, 373-377 (1999).
3. Wang, H., PhD thesis, Dept. of Phys. and Astron., UCLA (1998).
4. Benetti, P. et al., Nucl. Instrum. Methods A 329, 361 (1993).
5. Bottino, A. et al., Astropart. Phys. 2, 77-90 (1994).
6. Nath, P. and Arnowitt, R., Phys. Rev. Lett. 74, 4592-4595 (1995).